\documentclass[a4paper,11pt]{article}
\pdfoutput=1 
\usepackage{jheppub}
\usepackage{subcaption}
\usepackage[T1]{fontenc} 
\usepackage{siunitx}
\usepackage{placeins}

\newcommand{\myunit}[1]{$\, \mathrm{#1}$}

\def\GeV{\ifmmode {\mathrm{\ Ge\kern -0.1em V}}\else \textrm{Ge\kern -0.1em V}\fi}

\title{\boldmath Design, Construction and Performance Tests of a Prototype MicroMegas Chamber with Two Readout Planes in a Common Gas Volume}

\author[a]{Bernard Brickwedde}
\author[a]{Andreas D\"udder}
\author[a]{Matthias Schott}
\author[a]{Eda Yildirim}
\affiliation[a]{Johannes Gutenberg-University, Mainz, Germany}

\emailAdd{matthias.schott@cern.ch}

\abstract{
In this paper, the design and the performance of a prototype detector based on MicroMegas technology with two detection planes in a common gas volume is discussed. The detector is suited for the forward region of LHC detectors, addressing the high-rate environment and limited available space. Each detection plane has an active area of $9 \times 9$\myunit{cm^2} with a two-dimensional strip readout and is separated by a common gas region with a height of 14 mm. A micro-mesh, working as a cathode, is placed in the middle of the common gas volume separating it into two individual cells. This setup allows for an angle reconstruction of incoming particles with a precision of $\sim 2\,$mrad. Since this design reduces the impact of multiple scattering effects by the reduced material budget, possible applications for low energy beam experiments can be envisioned. The performance of the prototype detector has been tested with a 4.4\myunit{GeV} electron beam, provided by the test beam facility at DESY.}

\begin{document}

\maketitle

\section{Introduction}
\label{Sec:Intro}

In recent years, MicroMegas (\textbf{Micro}-\textbf{Me}sh \textbf{ga}seous \textbf{s}tructure) detectors  \cite{Giomataris:1995fq} received significant attention in the development of precision and cost-effective tracking detectors in nuclear and high energy physics experiments, e.g. they have been chosen as base-line technology for the upgrade project of the ATLAS muon system \cite{CERN-LHCC-2015-020,1748-0221-10-02-C02026}. A detailed introduction to the basic operation principle of a MicroMegas detector can be found in the literature \cite{Giomataris:1995fq, Alexopoulos:2010zz} and hence we give here only a brief summary. Standard MicroMegas detectors consist of two parallel-plates and a gaseous volume that is parted by a thin metallic mesh. The drift volume has a typical height of a few mm and is located between the upper plate and the mesh, while the amplification region has a typical height of 100$\,\si{\micro\meter}$ and is located between the separation mesh and the lower plate, which acts as readout plane. Electric fields are applied in the drift region and the amplification region and traversing charged particles ionize gas atoms in the drift region. The resulting electrons drift along the electric field lines towards the mesh, which appears transparent to these electrons when the appropriate voltages are applied. The electrical field in the amplification region is two orders of magnitude higher due to its small height. It is large enough to create an electron avalanche, leading to an amplification of the electron signal by a factor of $\sim10^4$ within less than a nanosecond. These secondary electrons together with the movement of the corresponding positive ions induce a signal in readout electrodes on the lower plate. The large electric fields in the amplification region can lead to sparks, resulting in a dead-time and potential damage to the detector and the subsequent front-end electronics. A resistive protection layer is therefore typically deposited on top of the readout strips \cite{Alexopoulos:2011zz}. 

Standard MicroMegas detectors achieve a spatial resolution in the order of $50-100\,\si{\micro\meter}$ and allow for a rough reconstruction of the incoming angle of traversing particles. One possible realization, using a single MicroMegas readout layer, is the so-called Micro-TPC method \cite{1748-0221-10-02-C02026}. However, its angular resolution is very limited, in particular for perpendicular incident particles. A pair of two MicroMegas detectors would allow for a precise angle measurement, however requires a relatively large volume and implies subsequently also a significantly larger material budget. In this paper, the design and the performance of a prototype detector with two detection planes in a common gas volume is discussed. This setup allows for a full angle reconstruction of incoming particles with reduced space requirements and material budget, compared to current detector designs. Possible applications are the high-rate and small available space environments in the forward region of LHC detectors, but also low energy beam experiments, where the effect of multiple scattering has to be reduced.

\section{Detector Layout and Construction}
\label{Sec:Detector}

\subsection{Detector Layout}

\begin{figure}[tb]
\begin{center}
\includegraphics[width=0.48\textwidth, height = 160px]{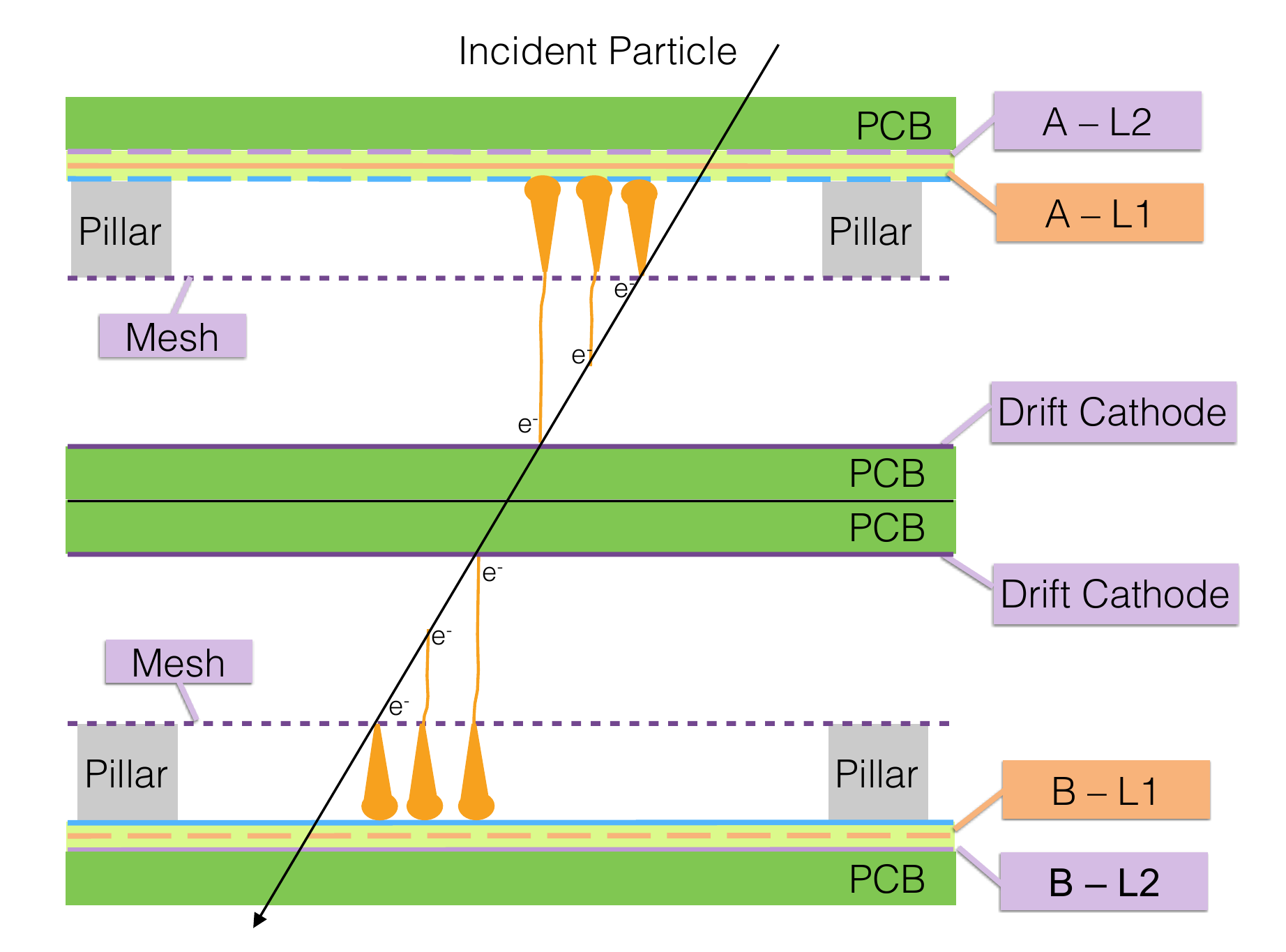}
\includegraphics[width=0.51\textwidth, height = 160px]{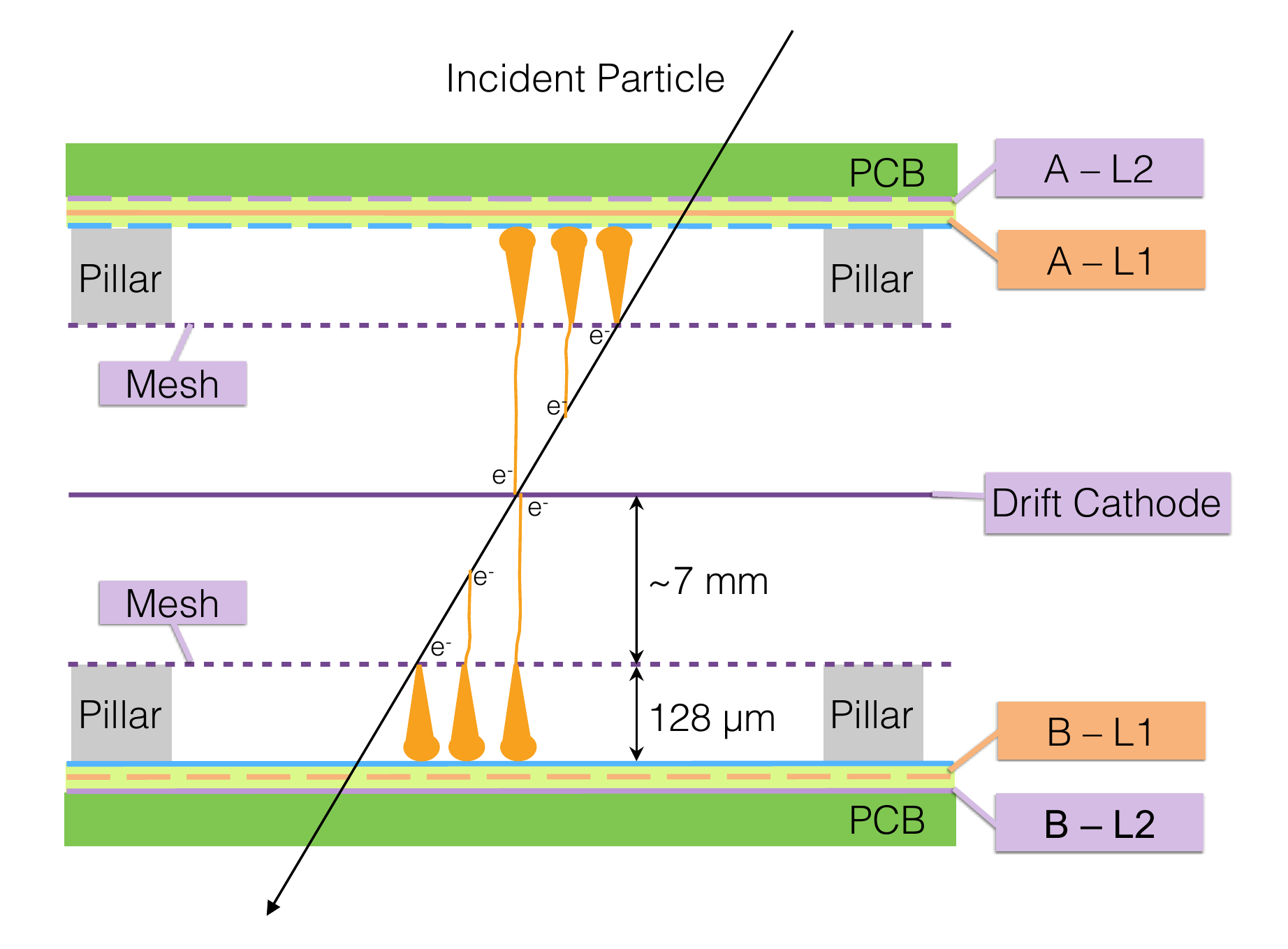}
\caption{Two MicroMegas detectors with a 2D readout structure with separated gas-volumes (left) and a common gas-volume (right) for the determination of incident angles of particles.}
\label{fig:ApproachComp}
\end{center}
\end{figure}

The measurement of two coordinates in two spatially separated planes is required in order to precisely reconstruct the incident angle of an incoming particle. Standard MicroMegas detectors allow the reconstruction of one coordinate, hence four MicroMegas detector layers have to be used for a precise angle measurement. However, in recent years,  MicroMegas detectors with resistive two-dimensional readout structures have been built and extensively tested \cite{Byszewski:2012zz, Lin2014281}. In contrast to conventional MicroMegas detectors, this layout provides two independent readout electrodes in orthogonal directions (denoted as L1 and L2 in the following), printed on the same PCB. Using this approach, the precise reconstruction of incident angles can be performed by using only two MicroMegas detectors, as illustrated in figure \ref{fig:ApproachComp}, reducing the material budget and space requirements already by a factor of two. The situation can be further improved by replacing the central PCB layers by a metallic mesh, that is acting as drift cathode and thus creating a common gas volume (figure \ref{fig:ApproachComp}), which is the basic idea of the prototype detector, described in the following. \\ \par

Both readout planes, labelled as A and B in the following, are made of a 1.8\,mm thick PCB and comprise 2D readout electrodes in two orthogonal layers, with 360 copper readout strips each with a strip pitch of $250\,\si{\micro\meter}$. The readout strips of the upper layer in both planes (defined as L1-layer) are printed directly on top of the PCB. They are covered by resistive strips with a resistivity of $\sim 100\,\mathrm{M\Omega/cm}$ separated from the readout strips by a thin isolation layer. The lower layer in both planes (defined as L2-layer) is separated from the upper layer by $100\,\si{\micro\meter}$ of FR4, i.e. the same material used as isolating material in the PCB. The readout strips of the L2-layer have a width of $200\,\si{\micro\meter}$ and are placed parallel to the resistive strips, while the strips in the L1-layer have a width of $80\,\si{\micro\meter}$ and are placed perpendicularly to the resistive strips. The larger width of the L2-layer readout strips partially compensates for their weaker capacitive coupling. \\ \par

A  woven mesh, made of stainless steel with a density of 157 lines/cm and a line diameter of $18\,\si{\micro\meter}$, is mounted with a distance of $128\,\si{\micro\meter}$ on top of the each readout plane, supported by isolating pillars of $0.4\,\mathrm{mm}$ diameter, placed along a regular matrix with $2.5\,$mm spacing in both directions. The same woven mesh is used as a cathode between the two readout planes with a distance of 7\,mm. The full MicroMegas doublet has therefore an active volume of (WxLxH) $90 \times 90 \times 14$\myunit{mm^3}. 

\subsection{Simulation}

The full detector design was simulated with \textsc{Geant4} \cite{Agostinelli2003250} to estimate the expected multiple scattering effects on the track reconstruction of traversing particles. The comparison of the average scattering angle after the full detector design for different incident particle energies is compared to a detector concept with two separated MicroMegas detectors using a 2D readout scheme. In this study a decrease of $\sim 30\,\%$ on the multiple scattering angle for our detector design is observed.

\begin{figure}[tb]
\begin{center}
\includegraphics[width=0.49\textwidth, height = 160px]{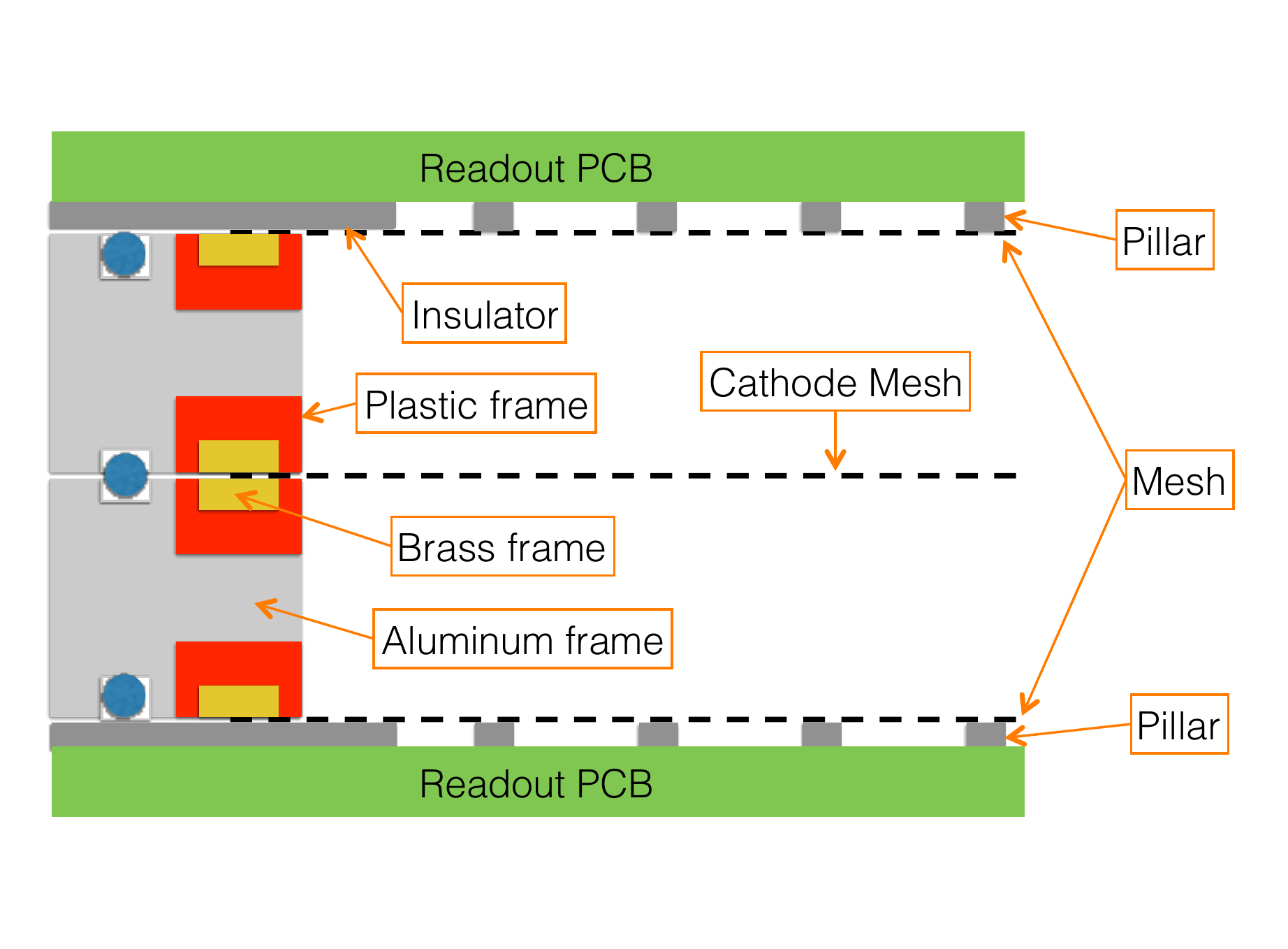}
\raisebox{28px}{\includegraphics[width=0.49\textwidth, height =100px]{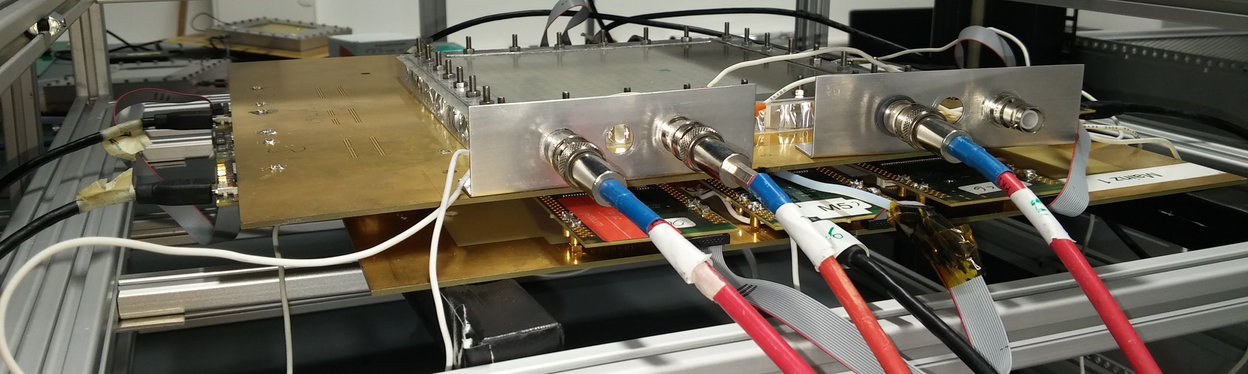}}
\caption{Schematic layout of the prototype detector (left) and a picture of the real detector (right).}
\label{fig:OverviewDetector}
\end{center}
\end{figure}

The expected angular resolution has been estimated using a \textsc{Garfield} \cite{Veenhof:1998tt} simulation of the detector, where the readout chain was independently simulated based on a dedicated electronic circuit model with \textsc{Qucs} \cite{Qucs}. An $Ar:C02$ (93:7) gas mixture and drift- and amplification-voltages of $500\,$V and $300\,$V have been used in the simulation. Additionally the deflection angle caused by multiple scattering as a function of the particle energy is simulated for electrons. As expected the deflection angle falls from 3 to less than 1$\,$mrad for electrons from 5 to 20$\,$GeV, respectively. 
\newpage

\begin{figure}[tb]
\begin{minipage}[hbt]{0.49\textwidth}
	\centering
	\includegraphics[width=0.99\textwidth,height = 160px]{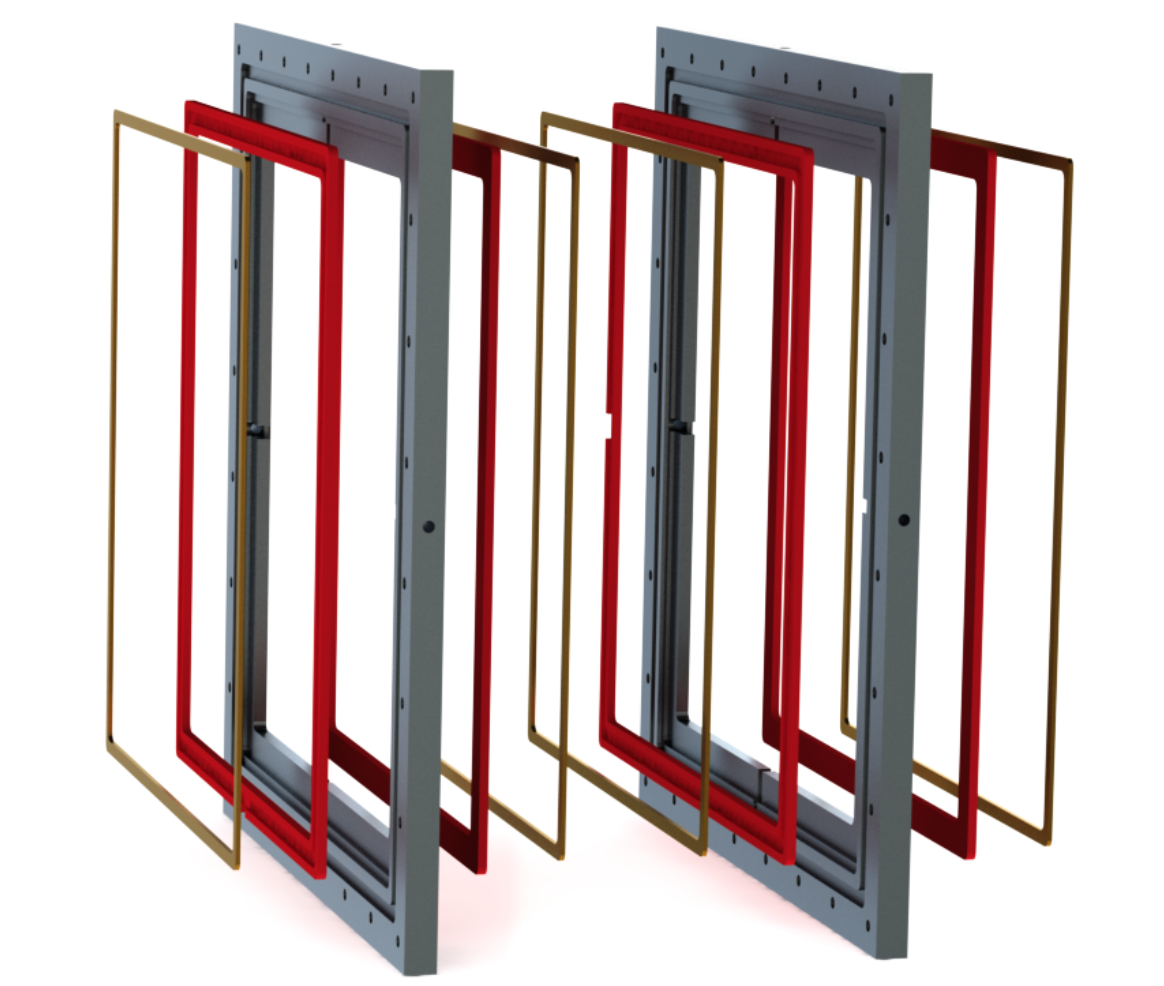}
\caption{\label{fig:consframe}View of the gas volume forming frames which contain the conducting loop for the mesh contact in an isolating plastic container (red).}
\end{minipage}
\hspace{0.1cm}
\begin{minipage}[hbt]{0.49\textwidth}
	\centering
	\raisebox{30px}{\includegraphics[width=0.99\textwidth, height = 130px]{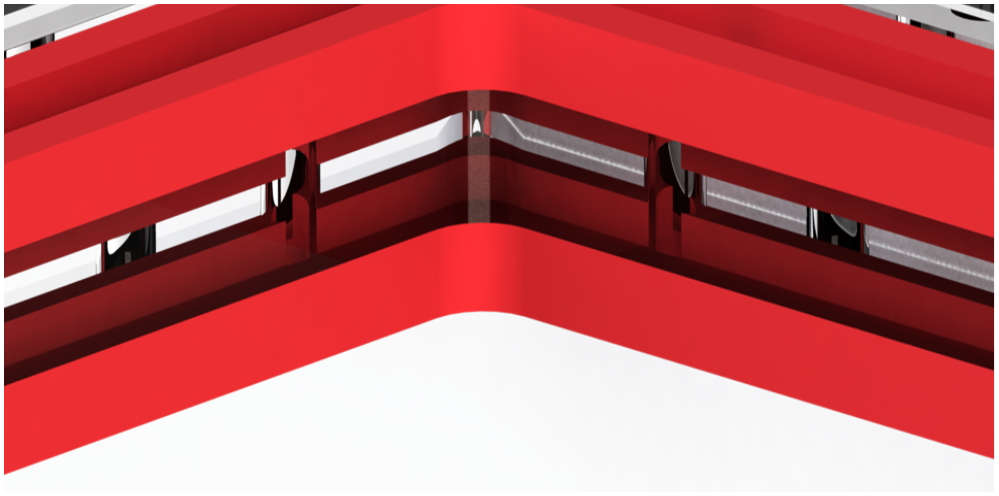}}
\caption{\label{fig:consplastic} One corner of the gas gap frame with the plastic isolation for the mesh contact. The drilled holes for the reinforcement plastic bolts are visible. }
\end{minipage}
\end{figure}

\subsection{Detector Construction}

The prototype detector comprises two gas frames as well as two readout printed circuit boards (PCBs) and is assembled as shown in figure \ref{fig:OverviewDetector}. The two 7$\,$mm high gas frames are made of aluminium and enclose the full gas volume. Additional plastic frames containing a conductor loop are glued to both the bottom and the top of each gas frame, illustrated in figure \ref{fig:consframe}. The intermediate plastic frames (figure \ref{fig:consplastic}) have been chosen to electrically isolate the gas frames from the meshes, which are stretched and glued on the plastic frames in a second step.  High voltage is supplied to the meshes by an external power supply using cables passed through the frames that are connected to the conductor loops. The cathode mesh is only glued to one frame, allowing to re-open the detector for maintenance and cleaning purposes. By pressing the second gas frame against the cathode mesh, it gets stabilized in the center of the common gas volume. In order to compensate for tensions or shearing forces in the amplification meshes and thus provide a uniform stretching, holes were drilled through both the gas- and plastic-frames and reinforced with plastic bolts. The gas frames and readout PCBs are arranged according to figure \ref{fig:OverviewDetector} and pressed together by screws. Note that due to manufacturing reasons the readout planes have been mounted with a rotation of 90$^\circ$ with respect to each other.

\par
In order to keep the system gas tight o-rings are placed between the gas frames as well as between the gas frame and the PCB on both sides. The detector can be stabilized by adding an additional layer of stiffening structures behind the PCBs. The reinforcement consist of two $0.5\,$\myunit{mm} FR-4 sheet enclosing a $10\,$\myunit{mm} high aluminium honeycomb.

\begin{figure}[tb]
\begin{center}
\includegraphics[width=0.49\textwidth, height=160px]{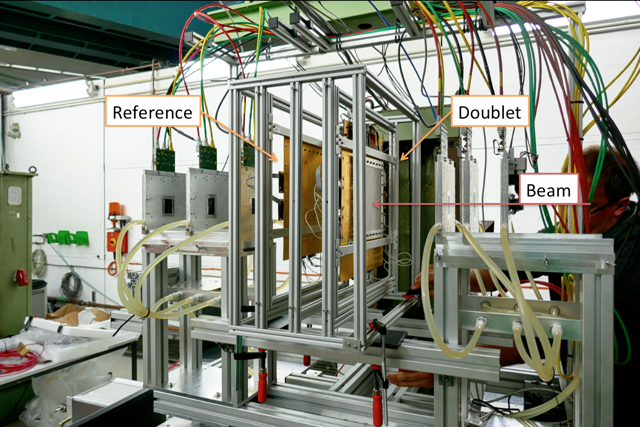}
\includegraphics[width=0.49\textwidth, height=160px]{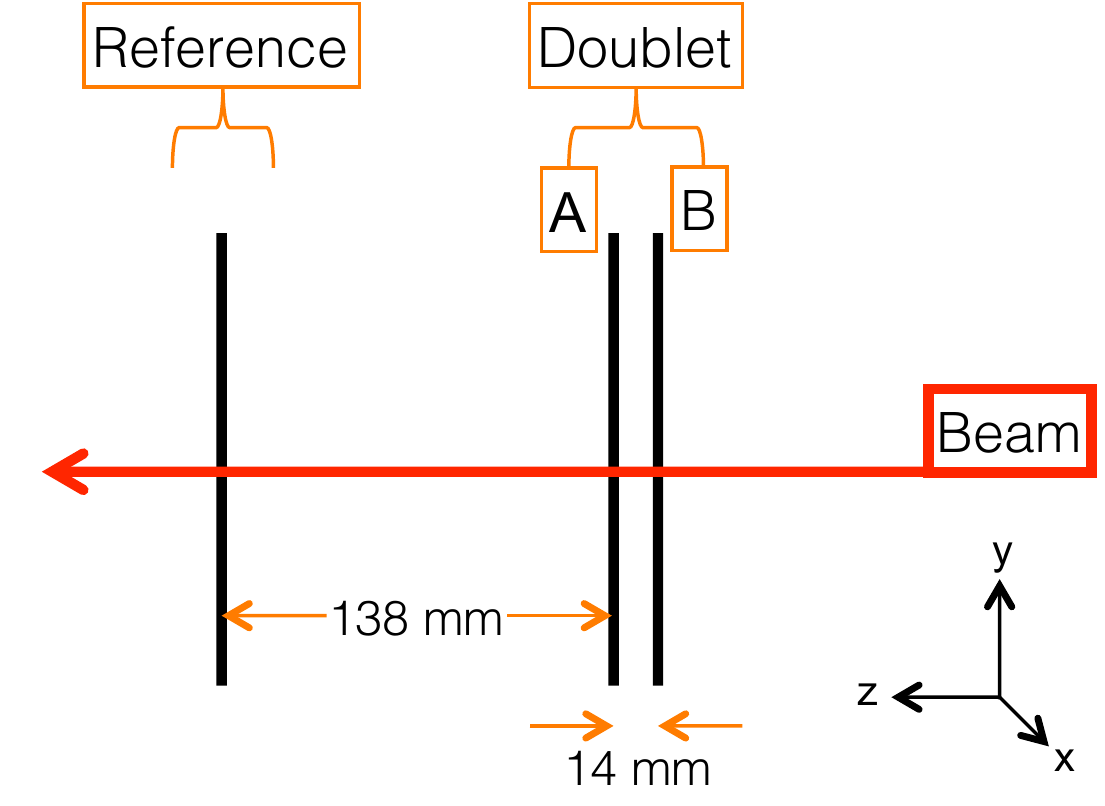}
\caption{Picture of the experimental setup at the DESY test beam facility (left) and the corresponding schematic layout (right).}
\label{fig:Experiment}
\end{center}
\end{figure}

\section{Performance Measurement}
\label{Sec:Performance}

\subsection{Experimental Setup and Signal Characteristics}

The performance of the prototype detector has been studied using data from a 4.4\myunit{GeV} electron beam at the DESY II test beam facility \cite{web-tb} with a 200\,Hz rate. In the following we will concentrate on the main performance parameters, such as efficiency, spatial and angular resolution. \\ \par

The experimental setup of the test-beam campaign is illustrated in figure~\ref{fig:Experiment}. The beam size is limited by using a 10x10\,mm collimator. The prototype doublet is placed 223\,mm after a scintillator trigger, followed by a second independent MicroMegas detector with a 2D readout scheme in 138\,mm distance that acts as reference chamber~\cite{Lin2014281}. Both detectors used an $Ar:CO_2$ (93:07 ) gas-mixture with drift- and amplification-voltages of 560\,V and 540\,V, respectively. The detector performance was measured for different tilt angles of the detector w.r.t. the incoming beam, ranging from $0^\circ$ (i.e. perpendicular incident beam) to $10^\circ$ around the vertical and horizontal axes, respectively. For each measurement, at least 100 000 events were recorded in the reference chamber. \\ \par

The detectors are read out through the RD51 Scalable Readout System (SRS)~\cite{Martoiu:2013aca,Martoiu:2011zja} and APV25 (Analog Pipeline Voltage chips with $0.25\,\si{\micro\meter}$ CMOS technology) \cite{APVJones} hybrid cards. The APV25 chip has 128 channels and delivers analog CR-RC shaped signals sampled at 40 MHz. Using HDMI cables, the analog signal from the APV25 is transmitted to the back-end board of the SRS DAQ where the signal is processed and sent to a PC to be saved for further analysis. Here, a zero suppression algorithm is applied. The chip was used in a mode that samples the integrated charge over up to 27 time bins of 25$\,$ns. \\ \par

For each recorded event in the reference chamber and the prototype detector,  the maximum of the integrated charge samples per channel, called signal charge is measured, can be extracted in units of ADC counts.

\begin{figure}[tb]
\begin{center}
\includegraphics[width=0.49\textwidth,height=160px]{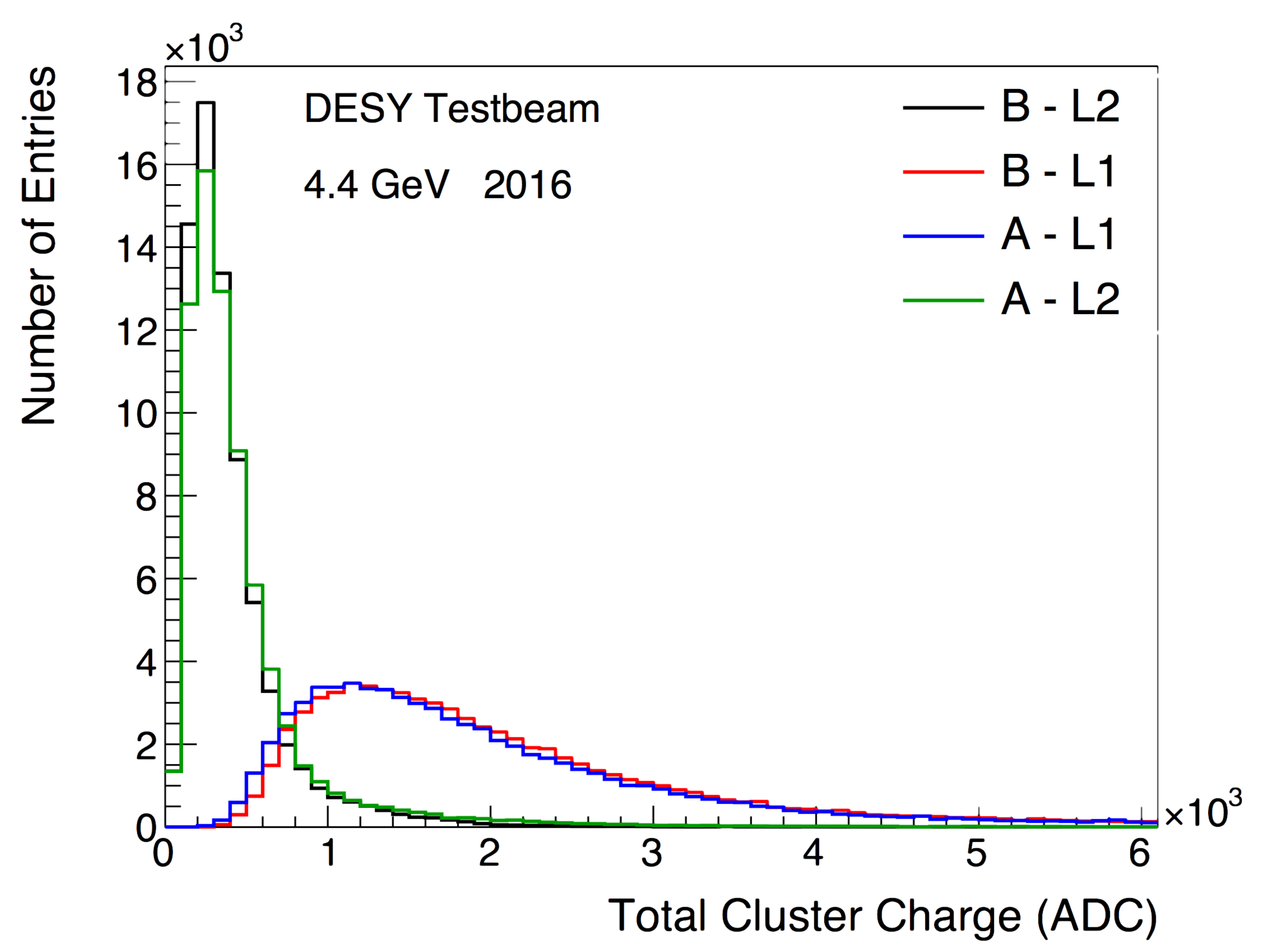}
\includegraphics[width=0.49\textwidth,height=160px]{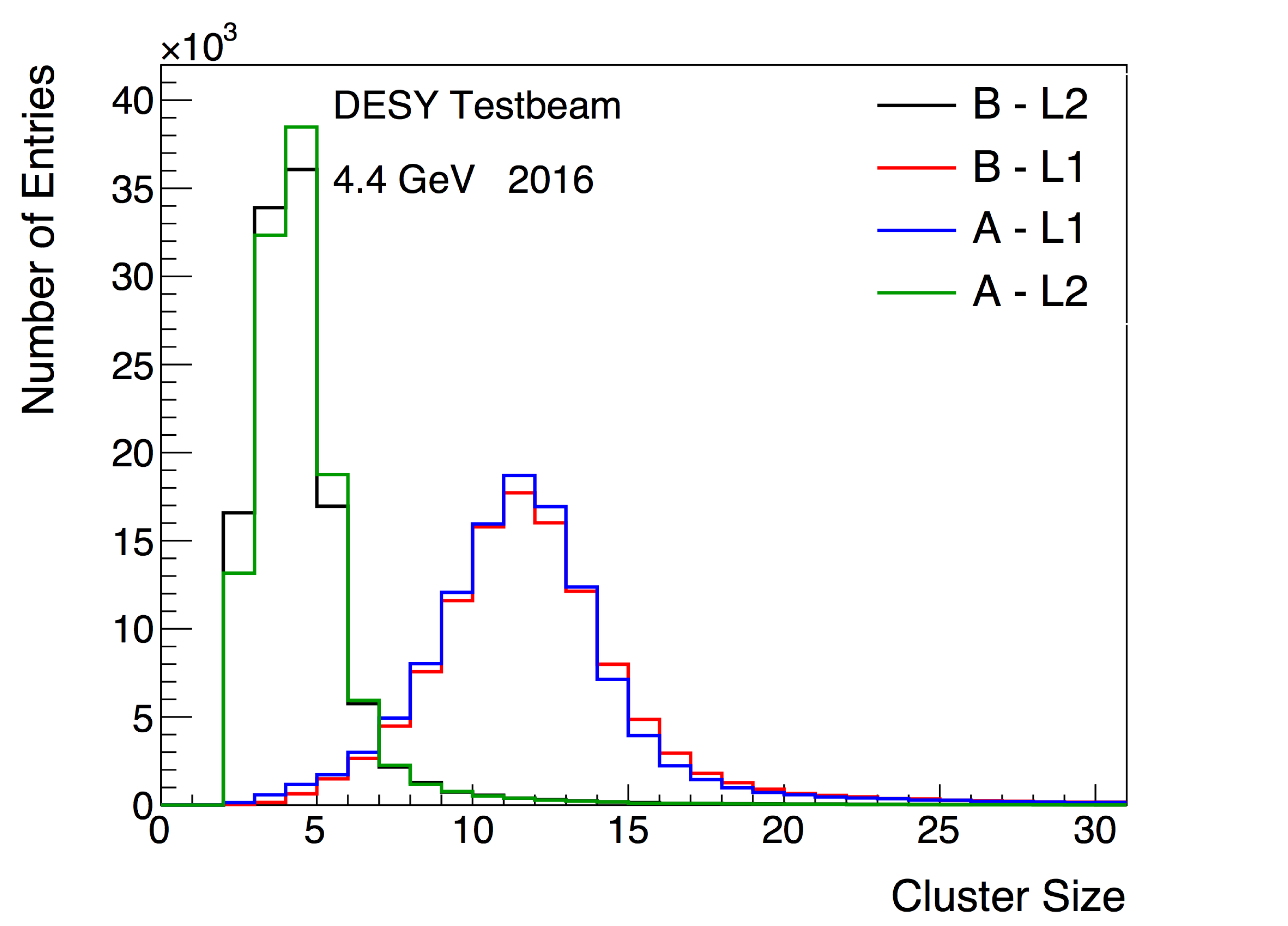}
\caption{Signal charge (left) and cluster size (right) for all layers of the prototype detector, measured at a perpendicular incident electron beam with an energy of $4.4\,$GeV.}
\label{fig:signalchar}
\end{center}
\end{figure}

\begin{figure}[tb]
\begin{center}
\includegraphics[width=0.49\textwidth, height = 160px]{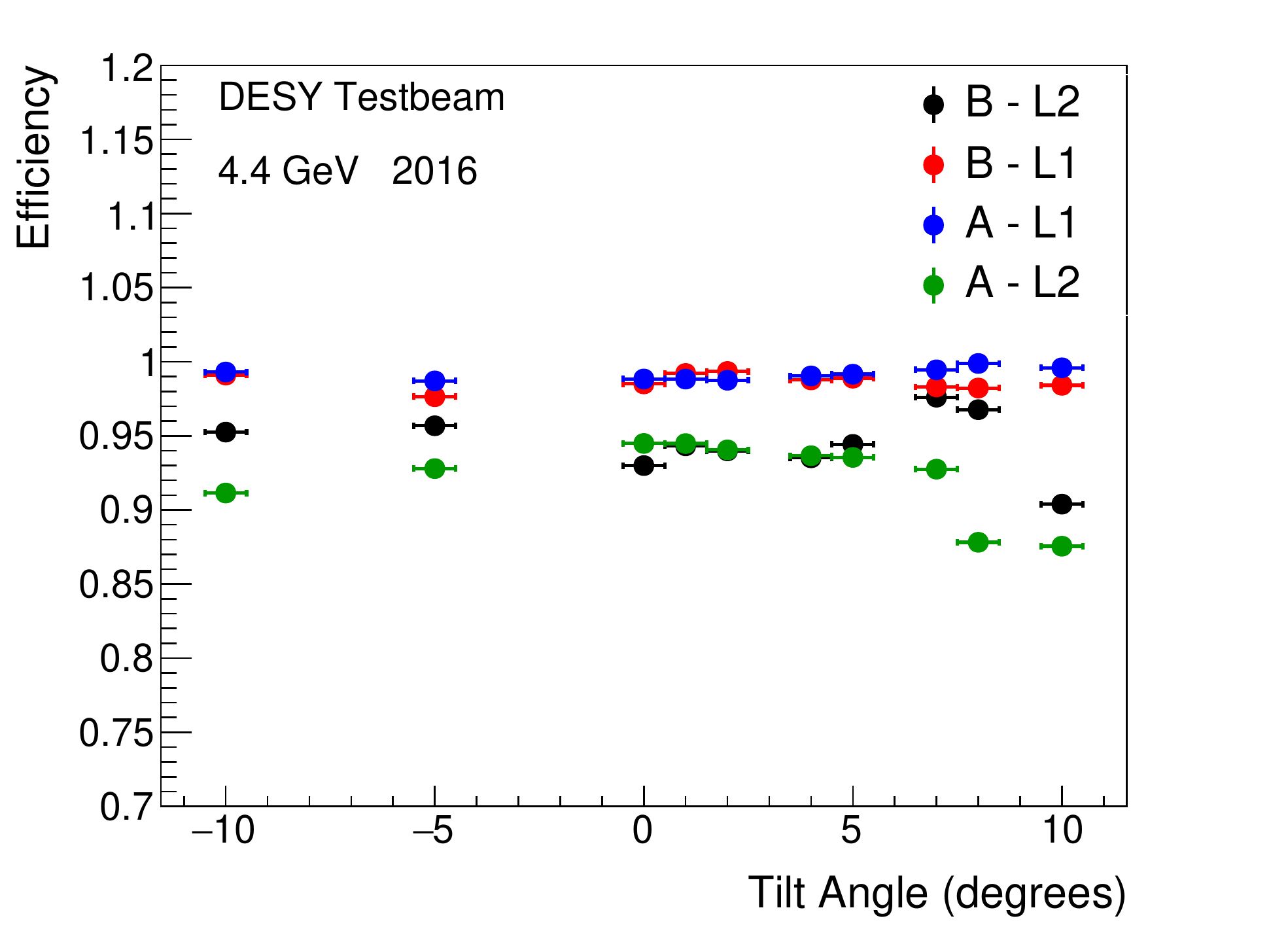}
\includegraphics[width=0.49\textwidth, height = 160px]{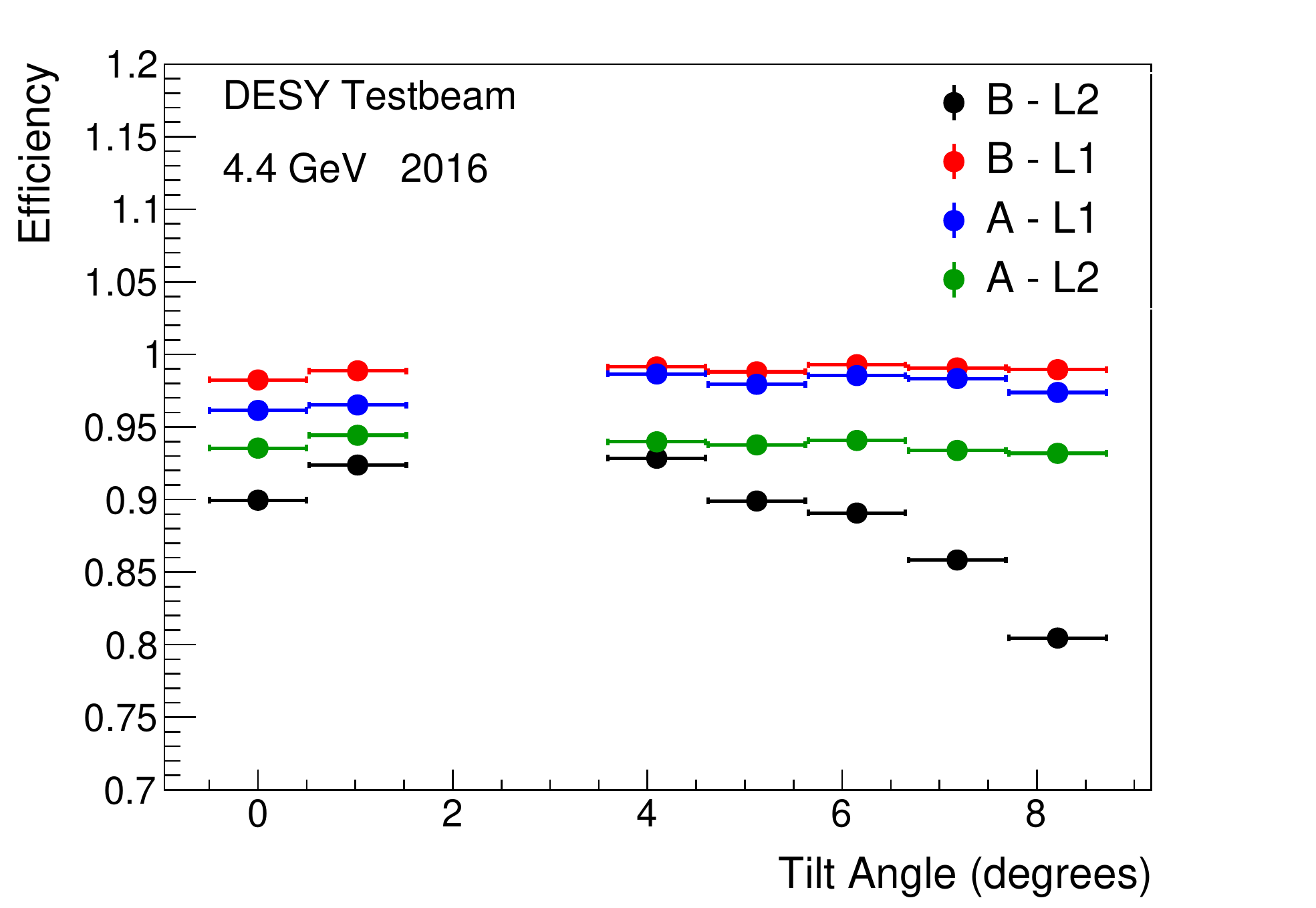}
\caption{Measurement of the cluster reconstruction efficiency for all layers of the prototype detector in dependence of the tilt around the vertical (left) and the horizontal axes (right).}
\label{fig:effTilt}
\end{center}
\end{figure}

Channels that have less than 20\myunit{ADC} counts are considered as noise. At least two neighboring channels with \myunit{ADC} count values above 20 are merged to clusters, where the cluster position is defined by the weighted mean of the integrated charge and the center of the corresponding readout strip. The cluster positions in the L1- and L2-readout layer are taken as an estimate of the hit-position of the traversing particle on the readout plane. The average signal charge as well as the cluster size, i.e. the number of merged neighboring channels, for all layers and different tilt angles is shown in figure \ref{fig:signalchar}. \\ \par

In general, the cluster size is 3-6 for the L2 layers and 6-10 for the L1 layers. This is due to the fact that the readout strips in L1 are perpendicular to the resistive strips, i.e. also perpendicular to the induced charges. Moreover twice as high integrated charge values for L1 compared to L2 are observed due to their smaller distance to the resistive layer. A detailed discussion of the signal characteristics of a MicroMegas detector with a similar 2D readout structure can be found in \cite{Lin2014281}.

\subsection{Efficiency Measurements}

Reconstructed clusters, caused by traversing particles, in the reference chamber as well as in the layers A-L1 and A-L2 of the prototype chamber can be used to extrapolate the expected hit position on the readout plane B. Similarly, the expected hit position on the readout plane A can be measured, using the reconstructed cluster information of the readout plane B. This extrapolation is the basis for the following discussion of efficiency and resolution measurements. 

A cluster is defined as reconstructed, if its measured position is within $1\,\mathrm{mm}$ of the expected position, defined by the track extrapolation of the remaining layers. No further requirements on the cluster characteristics itself are applied. The results for all layers of the prototype chamber for different tilt angles are shown in figure \ref{fig:effTilt}. While a uniform reconstruction efficiency above 98$\,$\% for all incident beam angles in the layers A-L1 and and B-L1 is observed, a clear degradation to 90$\,$\% is seen for layers A-L2 and B-L2, resulting from the larger distance between resistive strips and readout. Furthermore, a drop of the efficiency of layer B-L2 for larger tilt angles around the horizontal axis is observed. This can be explained by a localized inefficiency of several readout strips in layer B-L2, which are more affected by larger incident angles and therefore larger cluster sizes. 

\subsection{Spatial and Angular Resolution}
\label{sec-spa-res}

\begin{figure}[tb]
\begin{center}
\includegraphics[width=0.49\textwidth, height=160px]{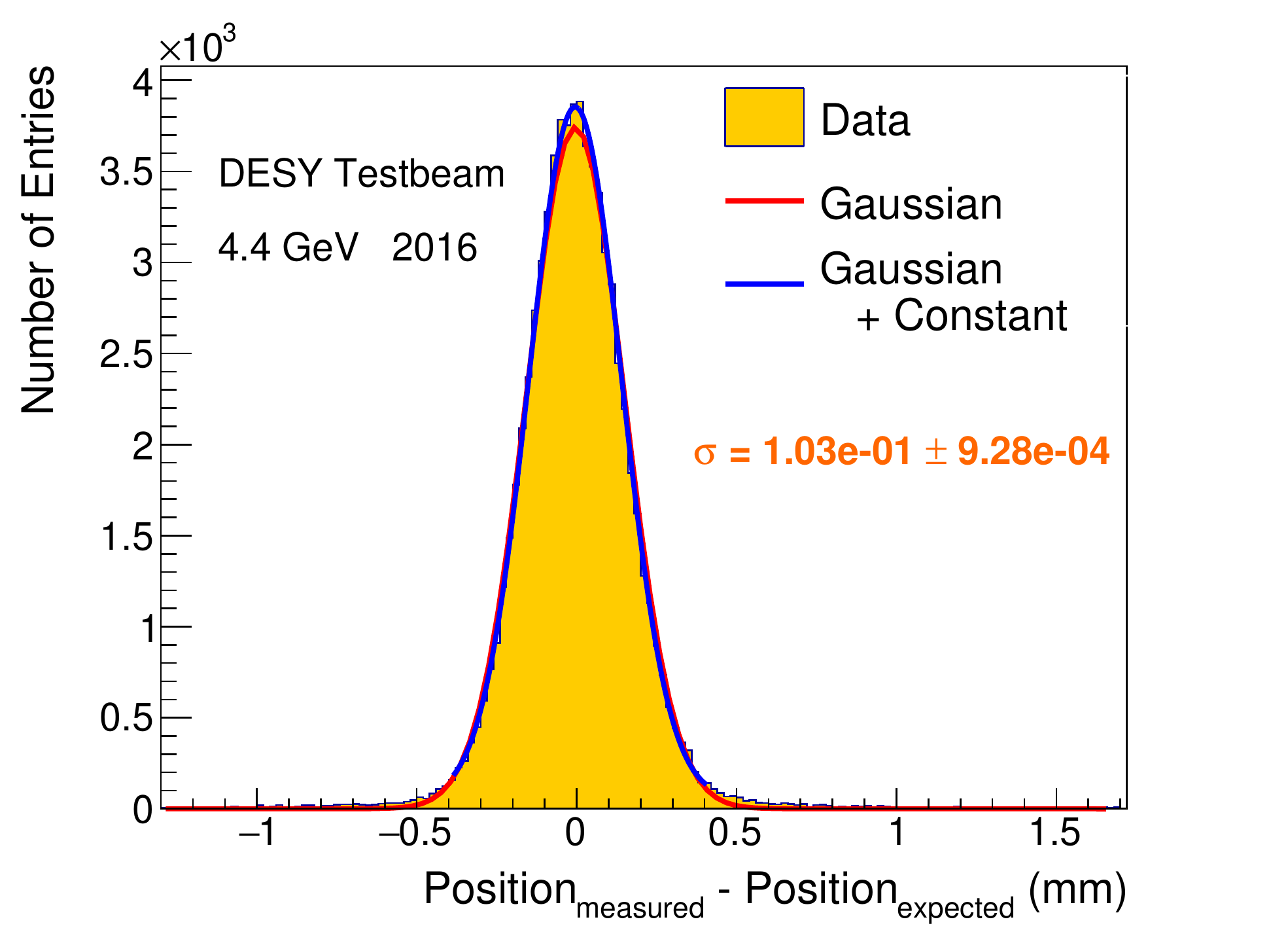}
\includegraphics[width=0.49\textwidth, height=160px]{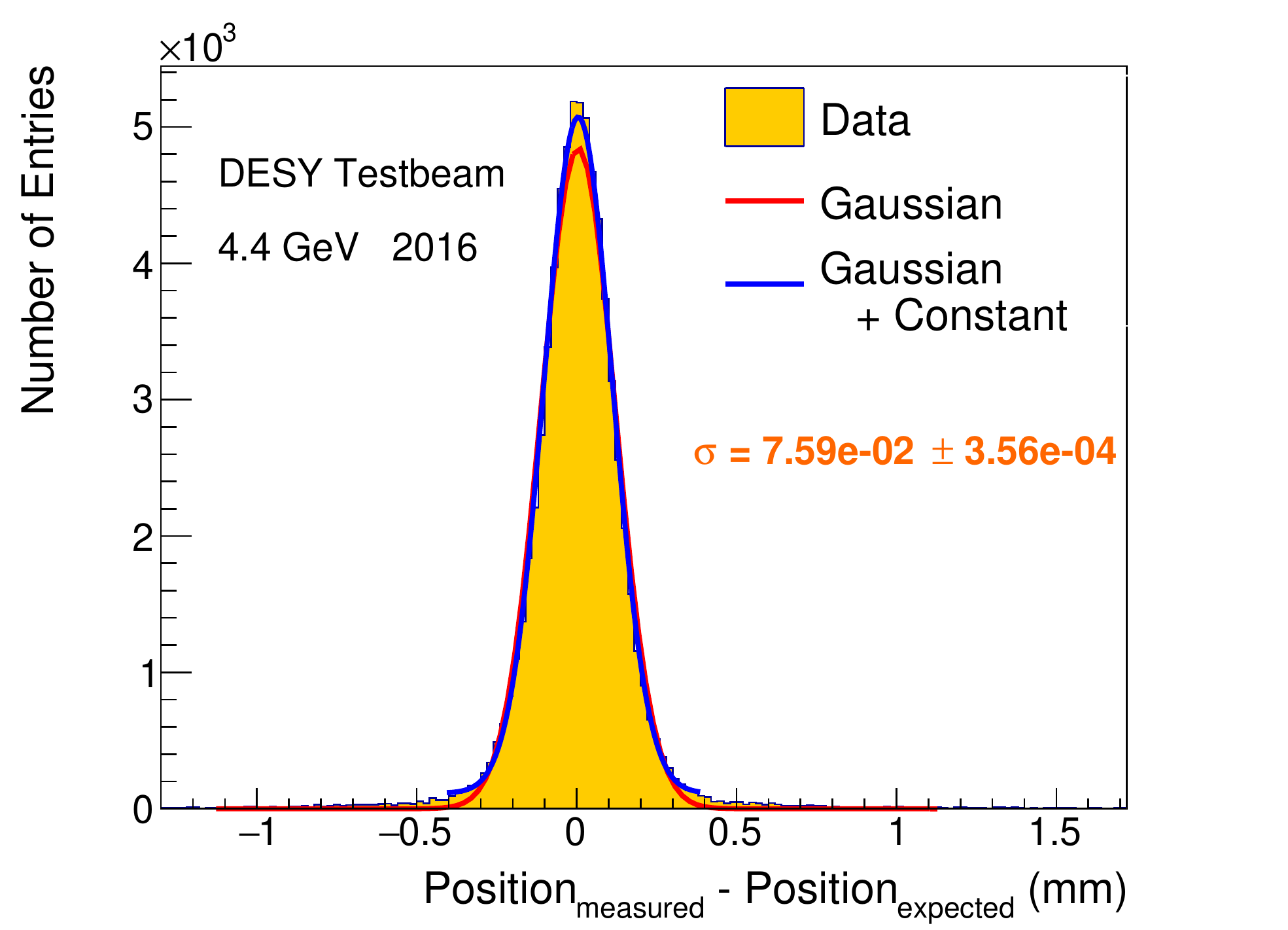}
\caption{Spatial residual distribution for layer B-L2 (left) and B-L1 (right) taken at $4^\circ$ of tilt around horizontal axis. }
\label{fig:spacialRes1}
\end{center}
\end{figure}

The spatial resolution is calculated for all recorded events containing one reconstructed cluster in each layer of the prototype chamber as well as in the reference chamber. The distributions of the residual differences between the expected position\footnote{based on the track extrapolation of the measurements at the reference chamber and readout layers A-L1 and A-L2} and the measured position for layer B-L1 and B-L2 is shown as  example in figure \ref{fig:spacialRes1}, where the detector was tilted by $4^\circ$ around the horizontal axis\footnote{The systematic uncertainty on the spatial resolution is estimated by comparing two functional fits to the residual distribution, once using a simple gaussian fit and once using a gaussian fit plus an offset parameter.}. Their widths are a measure of the spatial resolution, but have to be corrected for the geometrical effects and the  uncertainty of the extrapolated cluster position, which is given by the spatial resolutions of the remaining layer and reference chamber. \\ \par

\begin{figure}[tb]
\begin{center}
\includegraphics[width=0.49\textwidth, height=160px]{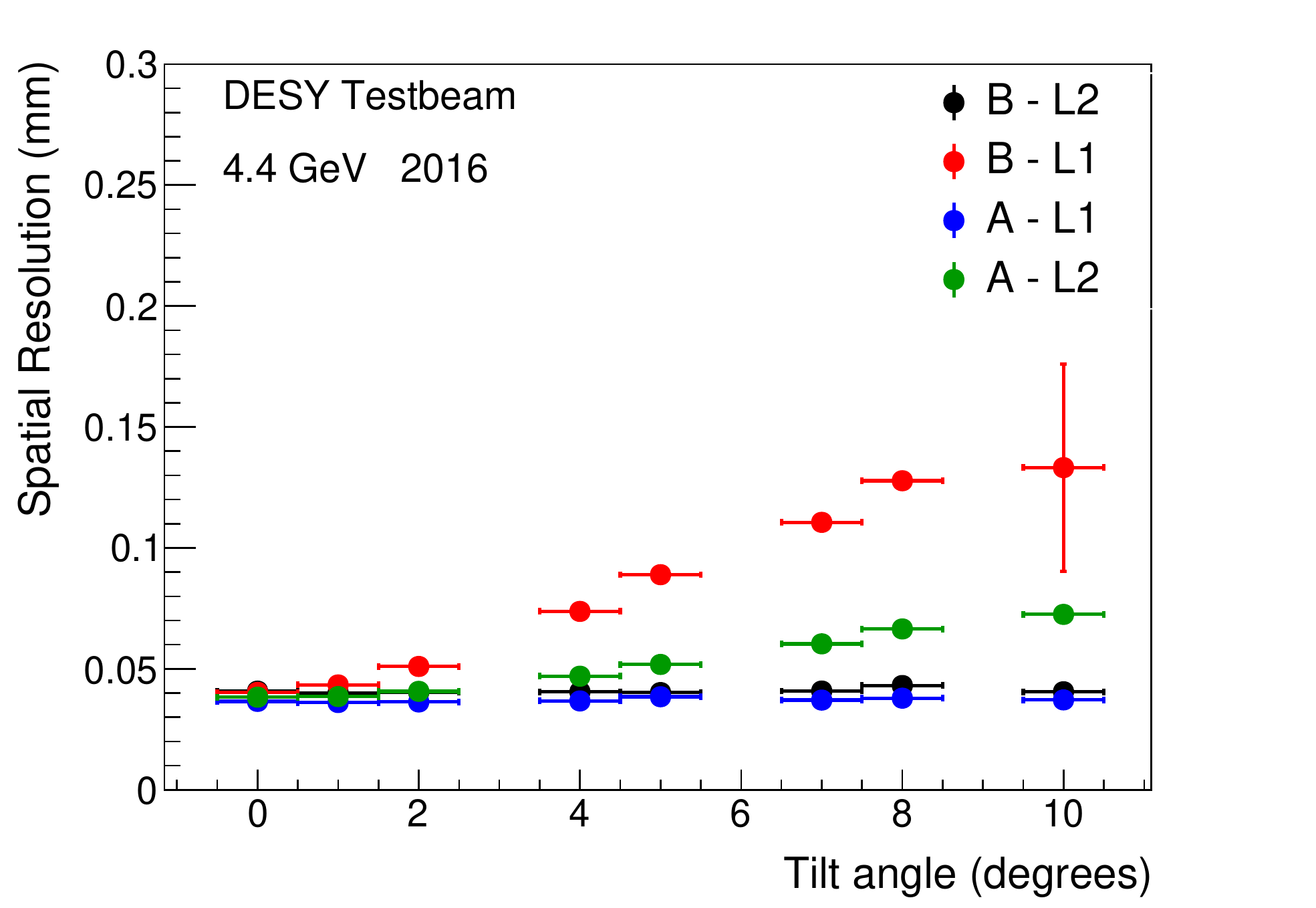}
\includegraphics[width=0.49\textwidth, height=160px]{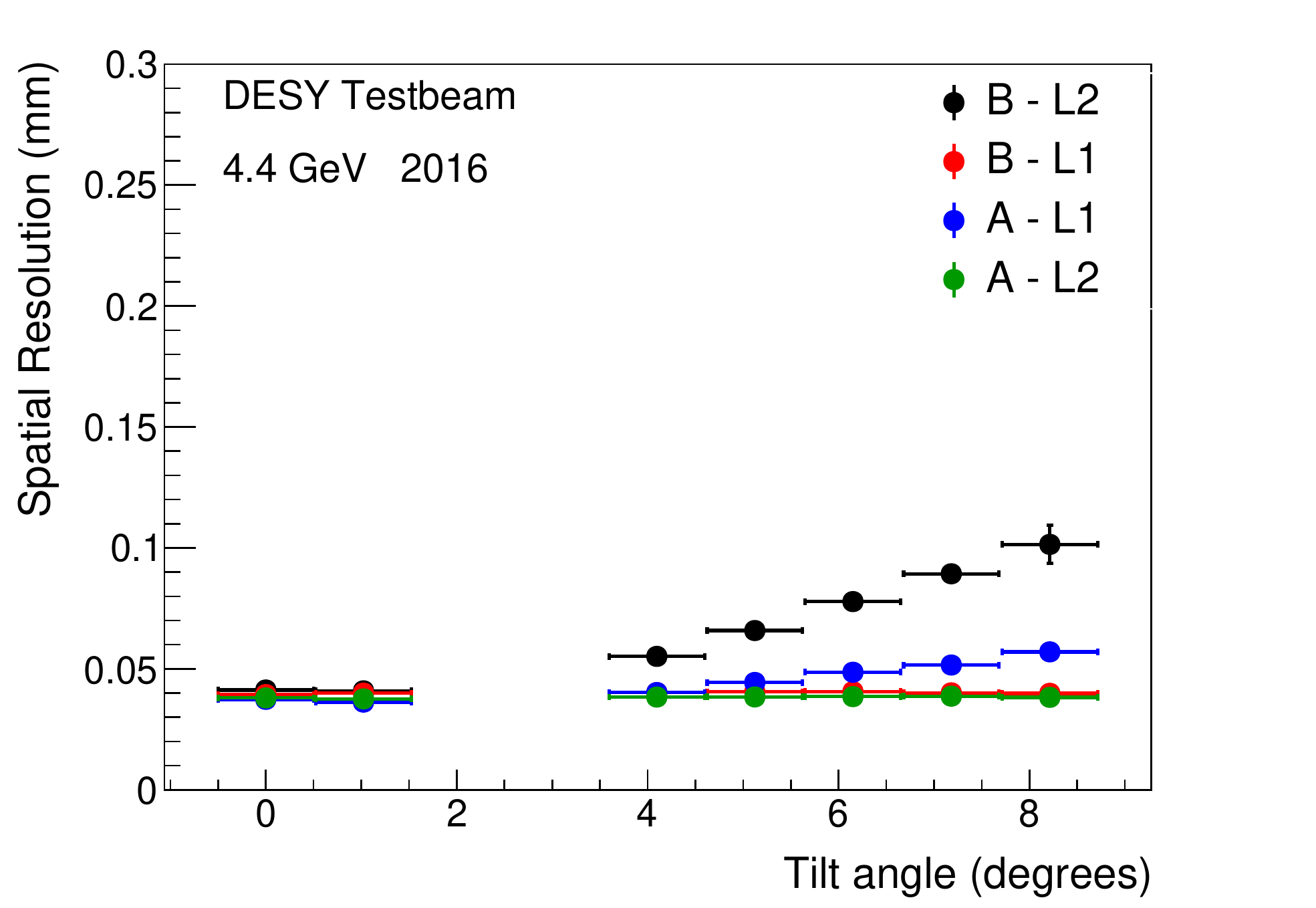}
\caption{Spatial resolution for all layers of the prototype chambers for tilts around the vertical (left) and the horizontal axes (right)}
\label{fig:spacialRes2}
\end{center}
\end{figure}

Due to the geometric setup of the prototype chamber and the reference chamber in the test beam, the uncertainty of the cluster position in the readout plane B is worse compared to plane A of the prototype detector  and the reference chamber. This effect must not be mistaken with the real detector resolution and can be corrected. The correction procedure has been derived using a detailed simulation of the test-beam setup, taking into account multiple scattering effects as well as the measured detector resolutions. The resulting spatial resolutions for different tilt angles are shown in figure \ref{fig:spacialRes2}. A resolution of 40 - 100$\,\si{\micro\meter}$ in both angular directions is observed. \\ \par

The resolution of the readout layers worsens with increasing tilt angle for a given rotation axis, as more drift electrons are created over a larger distance w.r.t. to the readout plane. This leads to wider induced signal, and hence a worse resolution\footnote{This effect can be in principle also corrected with an iterative approach, which was not used in this study}.

The difference in the rise of the resolution of the layers A and B is caused by a geometric effect and furthermore by the difference in the resolution of the prototype chamber and the reference chamber. The prototype chamber has a 2~mm larger drift region compared to the reference chamber. By this geometric change, an incoming particle with an e.g. $8^\circ$ incident angle passes a 30\% larger distance on the readout layer and thus leading to a wider electron spread, which is not corrected for in the final resolution evaluation in this analysis.

The angle of the incoming particle beam is measured by comparing the cluster positions in both readout planes A and B of the prototype detector. The distance between the two layers, $\Delta z$, in combination with the difference of the cluster positions $\Delta x$, where $\Delta x$ is the difference in the parallel coordinate in both readout planes, allows for the determination of the incident angle $\theta_x=\arctan(\Delta x/\Delta z)$. The angle resolution is therefore directly determined by the spatial resolution of the cluster position determination in both planes.

\FloatBarrier

\begin{figure}[t!b]
\begin{minipage}[t]{0.49\textwidth}
	\centering
	\includegraphics[width=0.99\textwidth, height=160px]{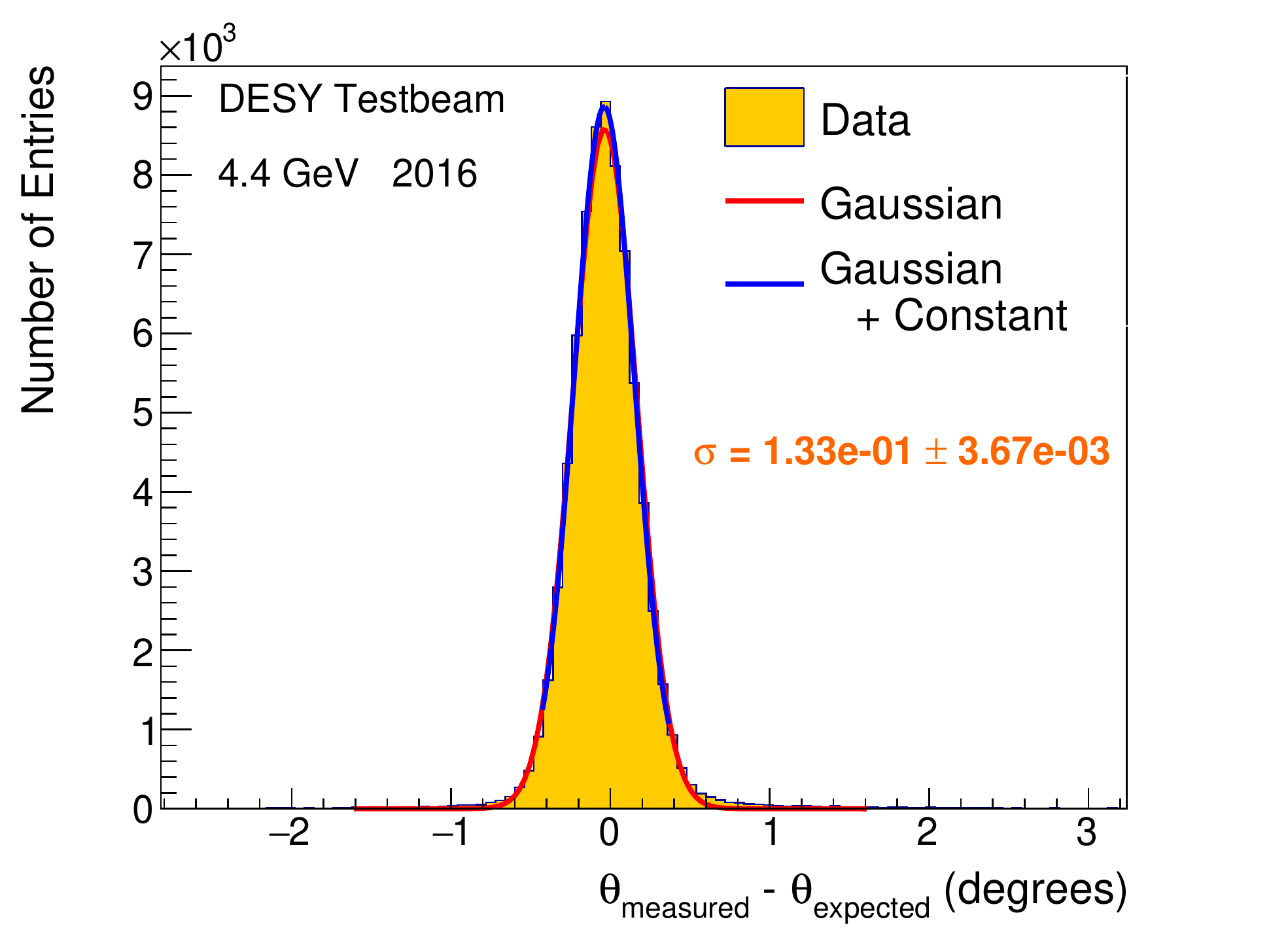}
\caption{\label{ang-residual}Angular residual distribution at 4 degrees of tilt around horizontal axis.}
\end{minipage}
\hspace{0.1cm}
\begin{minipage}[t]{0.49\textwidth}
	\centering
	\includegraphics[width=0.99\textwidth, height=160px]{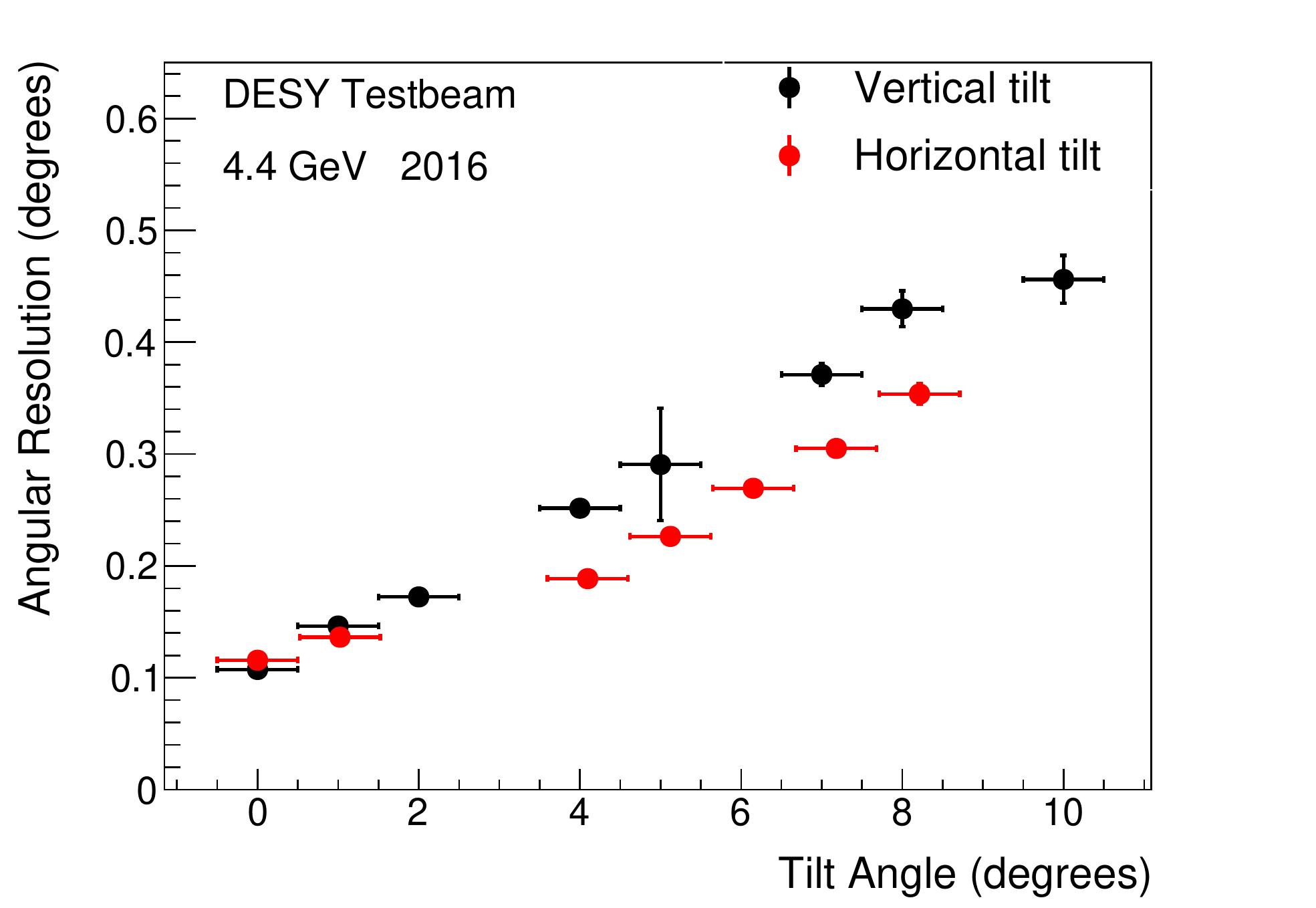}
\caption{\label{ang-resolution}Angular resolution dependence on tilt angle.}
\end{minipage}
\end{figure}

\FloatBarrier

Figure \ref{ang-residual} shows the angular residual distribution of the data taken at $4^\circ$ tilt angle around the horizontal axis. The gaussian width of the residual distribution is taken as measure of the angular resolution. Figure \ref{ang-resolution} shows the dependence of the angular resolution on the tilt angles around vertical and horizontal axes. For perpendicular incident beams, we observe an angular resolution of $0.11^\circ$, which increases to $0.45^\circ$ for tilt angles of $10^\circ$. This dependence is a direct result of the observed spatial resolution dependence on the incident beam angle.

\section{Conclusion}
\label{Sec:Conc}

A MicroMegas based prototype detector with two readout layers, each comprising a two-dimensional readout, in a common gas volume, has been constructed and tested. This design is particularly suited for high-rate environments with limited available space, such as is present in the forward region of LHC detectors. In addition, the detector design offers a reduced material budget compared to previous approaches and allows for a precise angle measurement of incoming ionizing particles, making it an interesting alternative for low energy experiments. The detector performance has been tested in a 4.4\myunit{GeV} electron beam, provided by the test beam facility at DESY. A spatial resolution of 55\,$\mu \mathrm{m}$ and an angular resolution of from $0.11^\circ$ for perpendicular incident beam particles to $0.45^\circ$ for incident angles of $10^\circ$ has been measured. The detection efficiency was found to be above $90\%$, limited by problems in the readout chain during the test-beam measurements. 

\section*{Acknowledgements}

We thank A. Sydorenko, S. Webb and T. Alexopoulos for the useful comments during the preparation of this paper. This work is supported by the Volkswagen Foundation and the German Research Foundation (DFG). The measurements leading to electron beam results have been performed at the Test Beam Facility at DESY Hamburg (Germany), a member of the Helmholtz Association (HGF).

\bibliography{MMTwoRegions}

\end{document}